\documentclass[runningheads]{llncs}
\usepackage{graphicx}
\usepackage{cite}
\usepackage{algorithmic}
\usepackage{textcomp}
\usepackage{soul}
\usepackage{amsmath,amssymb,amsfonts}
\usepackage{tikz}
\usetikzlibrary{intersections,calc,arrows}
\usepackage{xcolor}

\title{Chemical Property Prediction Under Experimental Biases}
\author{
Yang Liu
\and
Hisashi Kashima
}
\institute{
Department of Intelligence Science and Technology, Kyoto University
}

\begin{document}

\maketitle

\begin{abstract}
Predicting the chemical properties of compounds is crucial in discovering novel materials and drugs with specific desired  characteristics. Recent significant advances in machine learning technologies have enabled automatic predictive modeling from past experimental data reported in the literature.
However, these datasets are often biased because of various reasons, such as experimental plans and publication decisions, and the prediction models trained using such biased datasets often suffer from over-fitting to the biased distributions and perform poorly on subsequent uses.
Hence, this study focused on mitigating bias in the experimental datasets.
We adopted two techniques from causal inference combined with graph neural networks that can represent molecular structures.
The experimental results in four possible bias scenarios indicated that the inverse propensity scoring-based method and the counter-factual regression-based method made solid improvements.
\end{abstract}

Predicting the chemical properties of compounds is crucial in discovering  novel materials and drugs with specific desired characteristics.
Various computational approaches, including those based on density functional theory, have been widely used to accelerate the discovery process; however, they remain expensive and time consuming.
In recent decades, researchers have shifted to data-driven approaches by fast-progressing machine learning technologies~\cite{rupp2012fast}.
Recently, this trend has been further accelerated by the remarkable development of deep learning; in particular, graph neural networks (GNNs) have achieved remarkable performance in predicting chemical properties via automatic feature extraction from molecular structures represented as graphs~\cite{duvenaud2015convolutional,gilmer2017neural}. Their applications have expanded to various tasks, such as molecular generation~
\cite{liu2018constrained,you2018graph,de2018molgan}, molecular explanation~\cite{ying2019gnnexplainer,akita2018bayesgrad}, and analysis of inter-molecular interactions~\cite{harada2020dual,Wang2020GoGNNGO}.

Accurate predictive models often rely on large-scale labeled datasets; they are frequently collections of knowledge (e.g., experimental results reported on scientific papers) that are the product of extensive scientific efforts.
Unsurprisingly, scientists do not uniformly sample molecules from a large chemical space at random nor based on their natural distribution. Rather, their decisions on experimental plans or publication of results are biased due to physical, economic, or scientific reasons.
For instance, a large proportion of molecules are not investigated experimentally because of molecular mechanics-related factors, such as solubility~\cite{llinas2007diclofenac}, weights~\cite{raymer2018lead}, toxicity~\cite{hann2011molecular},  and side effects, or molecular structure-related factors, such as  crystals~\cite{jia2019anthropogenic}.
Further, concerns about the cost and availability of molecules can be some reasons to exclude certain groups of molecules.
Conversely, popularity considerations based on current research trends~\cite{hattori2008predicting} and the experimental methods wherein each lab specializes~\cite{10.1039/9781847559388} influence the selection of compounds.
These propensities related to researchers' experience and knowledge can contribute to more efficient search and discovery in the chemical space; however, they influence the data in an undesirable manner.
Biases from human scientific research result in datasets that are sampled from distributions that differ from the natural ones.
Thus, prediction models trained using such biased datasets suffer from over-fitting to the biased distributions, leading to poor performance on subsequent uses~\cite{kearnes2016modeling,wallach2018most,chen2019alchemy}.

\begin{figure*}[t]
    \centering
    \includegraphics[width=\linewidth]{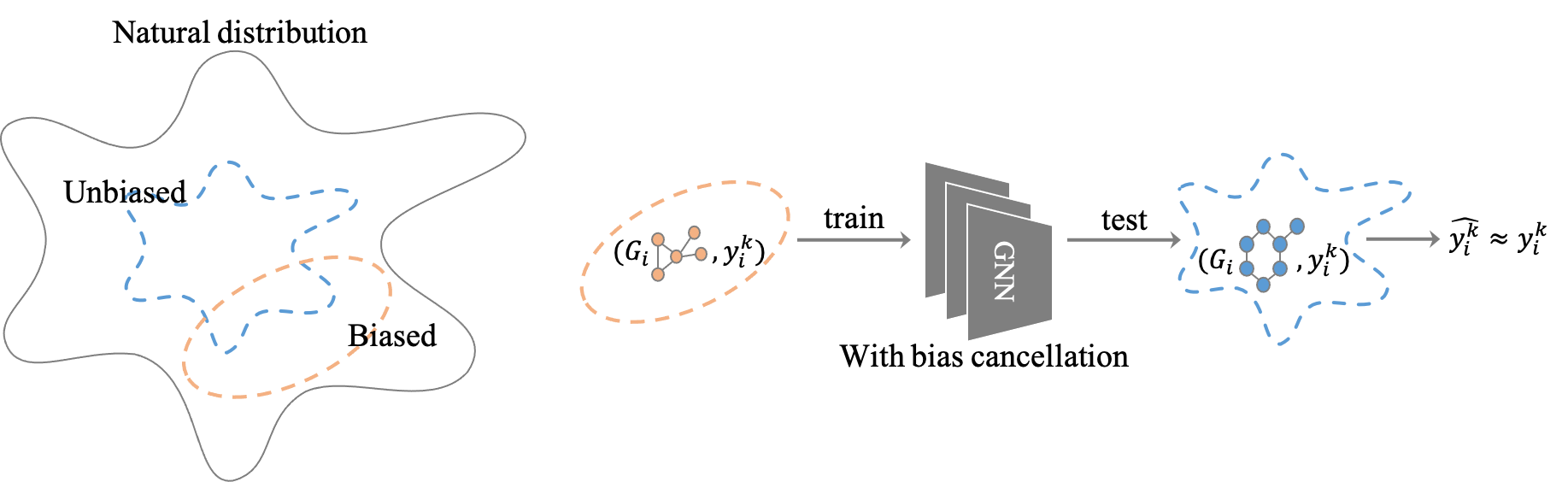}
    \caption{Main aim of our study. The left-hand figure shows the biased and unbiased distribution compared with the natural universe distribution of a chemical domain or sub-domain. The right-hand figure shows the proposed methods to train GNNs for chemical property prediction. Our aim was to apply bias cancelling techniques for GNNs to achieve significantly lower errors (i.e., MAE) when tested on a randomly sampled test dataset, whose distribution is similar to nature. $G$ is the molecular graph, and $y^k$ is the value of the $k$-th chemical property.}
    \label{fig:diagram}
\end{figure*}

Evidence on the existence of bias in scientific experiments and their harmful effects has been reported. However, almost no attempts to address this challenge have been found. In this study, we focused on mitigating the sources of bias in experimental datasets by applying two techniques from causal inference  (i.e., inverse propensity scoring (IPS)~\cite{imbens2015causal,Schnabel2016RecommendationsAT} and counter factual regression (CFR)~\cite{shalit2017estimating,NIPS2018_7529,hassanpour2019counterfactual}
).
With the IPS approach, we first estimate the propensity score function, which represents the probability of each molecule to be analysed, and then estimate the chemical property prediction model by weighting the objective function with the inverse of the propensity score.
The CFR approach is based on more recent advancements in the representation learning of deep neural networks for causal inference.
It consists of one feature extractor, several treatment outcome predictors and one internal probability metric, where the feature extractor obtains features that aid the different treatment outcome predictors and the internal probability metric, and the entire network is optimized in an end-to-end manner. Hassanpour et al.~\cite{hassanpour2019counterfactual} further introduce an importance sampling weight estimator to improve the CFR architecture.

Both approaches were implemented over a GNN to study the molecular structures of the compounds.
To the best of our knowledge, we are the first to combine chemical property prediction based on graph deep learning and sampling bias correction techniques.
Figure~{\ref{fig:diagram}} shows the main aim of this study.

Our experiments used two well-known large-scale dataset QM9~\cite{ramakrishnan2014quantum} and  ZINC~\cite{gomez2018automatic},
and two relatively smaller datasets ESOL and FreeSolv~\cite{wu2018moleculenet}.
QM9 consists of exhaustively enumerated small-molecule structures associated with 12 fundamental chemical properties~\cite{ruddigkeit2012enumeration},
In contrast, molecule structures in ZINC, ESOL, and FreeSolv are far less than exhaustively enumerated and they are associated with one property for each dataset.
Because determining how a publicly available dataset is truly affected by bias is impossible, we simulated four practical biased sampling scenarios from the dataset, which introduced significant biases in the observed molecules.
Under each biased sampling scenario, we validated our proposed models in predicting 15 chemical properties using 15 regression problems.
The experimental results indicated that both the two-step IPS approach and the more modern end-to-end CFR approach improved the predictive performance in all the scenarios on most of the targets with statistical significance compared with the baseline method
, and in addition, the CFR approach outperformed the IPS approach on most of the targets.

In summary, the main contributions of this paper are as follows:
\begin{itemize}
    \item We first address the problem of predicting properties of chemical compounds under experimental biases.
    \item We introduce two bias mitigation techniques, IPS and CFR, combined with GNN-based prediction of chemical properties.
    \item We validated the two proposed approaches using various experimentally biased sampling scenarios and demonstrated that both of them improves the predictive performance significantly.
\end{itemize}

\section{Related Work} 

The literature in the fields of chemical, physical, biological, and pharmaceutical sciences presents the existence of experimental biases. 
For instance, the types of compounds to be investigated are encouraged/discouraged by various physical and chemical properties of compounds, such as solubility~\cite{llinas2007diclofenac}, weight~\cite{raymer2018lead}, toxicity~\cite{hann2011molecular}, and side effects, or by factors related to the molecular structure, such as crystals~\cite{jia2019anthropogenic}.
In the pharmaceutical domain, drug likeness is an important factor for target selection, as exemplified by ``Lipinski's rule of five"~\cite{lipinski2004lead}.
Other sources of bias in the selection are concerns about the cost, availability, experimental methods~\cite{10.1039/9781847559388}, and  research trends~\cite{hattori2008predicting}.
This study focused on properties of compounds themselves; however, experimental biases in their interactions such as drug-target interactions have also been  reported~\cite{yildirim2007drug,mestres2008data}.
Although the literature indicates negative effects on the performance of predictive modeling using biased datasets~\cite{kearnes2016modeling,wallach2018most,chen2019alchemy}, to the best of our knowledge, no attempt has been made to mitigate the biases to improve the predictive performance.

Recent significant developments in deep neural networks have expanded their scope from vector data to texts and images, as well as to graph-structured data.
GNNs are actively being studied~\cite{Hamilton2017RepresentationLO,wu2020comprehensive} and successfully applied to chemical and physical domains~\cite{duvenaud2015convolutional,kearnes2016molecular,velivckovic2018graph,hamilton2017inductive,li2018adaptive,yan2018spatial,yu2017spatio}.
Specifically, we used one of the well-known fairly general GNNs~\cite{gilmer2017neural} in this study. However, more modern architectures have been proposed (e.g., one incorporating attention~\cite{velivckovic2018graph} and another with more expressive power~\cite{xu2018powerful}).
Most of previous studies assumed unbiased datasets, and this study is the first to address the bias mitigation problem in learning GNNs, except for the one considering classification \emph{on a single large graph}~\cite{guo2020learning}, while we considered the classification \emph{of a set of small graphs}.

The problems of learning prediction models, when the distributions of the training and test datasets are different, are called domain adaptation, covariate shift adaptation~\cite{quionero2009dataset}, or transfer learning~\cite{pan2009survey}, and they have been a major topic in machine learning.
Recently, deep neural networks for domain adaptation based on the concept of 
domain-invariant representation learning
were proposed~\cite{ganin2016domain,tzeng2017adversarial, tang2020discriminative, tanwani2020domain, long2017conditional, lee2019drop, ma2019gcan}, which have been primarily applied in the study of images.
The problem of biased observations is also discussed in the context of causal inference.
Inverse propensity scoring (IPS) is a general method from causal inference~\cite{imbens2015causal,Schnabel2016RecommendationsAT}, which has been successfully applied to various applications such as recommender systems~\cite{Schnabel2016RecommendationsAT, ma2019missing}, natural language processing~\cite{zhang2019selection}, and treatment estimation in different fields ranging from healthcare~\cite{eichler2016threshold}, economy~\cite{lalonde1986evaluating}, and education~\cite{zhao2017estimating}.
Meanwhile, the domain-invariant representation learning concept is also introduced in the context of causal inference~\cite{shalit2017estimating}.
Counter factual regression (CFR) is a classic method which has been successfully applied for predicting individual treatment effect~\cite{shalit2017estimating,NIPS2018_7529,hassanpour2019counterfactual} by obtaining balanced representation such that the induced treated and control distributions look similar.

\begin{table}[tb]
    \centering
    \caption{List of symbols used in this paper.}
    \begin{tabular}{l|l}
      Symbol in PROBLEM SETTING & Description \\
      \hline
      $\mathcal{G}$ & universe of molecules \\
      $\mathcal{D}^\text{train} = \{(G_i,y_i){\}_{i=1}^N}\subset \mathcal{G}$ & training dataset of $N$ molecules \\
        $\mathcal{D}^\text{test} = \{G_i{\}}_{i=N+1}^{N+M}$ & test dataset of $M$ molecules \\
      \hline
      $G_i  =(\mathcal{V}_i,\mathcal{E}_i, \sigma_i)\in \mathcal{G}$ & molecular graph \\
      $\mathcal{V}_i$ & set of graph nodes of $G_i$ \\
      $\mathcal{E}_i \subseteq \mathcal{V}_i \times \mathcal{V}_i $& set of edges of $G_i$ \\      
      $\sigma: \mathcal{V}_i \cup \mathcal{E}_i \rightarrow \mathcal{L}$ & node and edge label function \\
      $\mathcal{L}$ & set of node and edge labels \\
      $y_i \in \mathbb{R}$ & target chemical property value~~~~~~~ \\     
      \hline
    
    \end{tabular}
\\
\vspace{5mm}
    \begin{tabular}{l|l}
      Symbol in METHODS~~~~~~~~~~~~~~~~~ & Description \\
      \hline
      $\mathbf{m}_v^t \in \mathbb{R}^D$ & message of node $v$ in layer $t$ \\
      $\mathbf{h}_v^t \in \mathbb{R}^D$ & feature vector of node $v$ in layer $t$ \\      $d_i \in \{0,1\}$ & domain of $G_i$ \\
      $m_t: (\mathbf{h}_v^t,\mathbf{h}_u^t,\sigma(u,v))\rightarrow\mathbb{R}^D$ & GNN message function \\
      $a: \mathbb{R}^D\rightarrow\mathbb{R}^D$ & GNN activation function \\
      $u_t: (\mathbf{h}_v^t,\mathbf{m}_v^t)\rightarrow\mathbb{R}^D$ & GNN update function \\
      $r: \{\mathbf{h}_v^T\}\rightarrow\mathbb{R}^D$ & GNN graph-level readout function \\
      $f: \mathcal{G}\rightarrow\mathbb{R}$ & property predictor \\
      $f_{\text F}: \mathcal{G}\rightarrow\mathbb{R}^D$ & feature extractor \\
      $f_{\text L}: \mathbb{R}^D\rightarrow\mathbb{R}$ & label predictor \\
      $f_{\text W}: \mathbb{R}^D\rightarrow\mathbb{R}^2$ & weight estimator \\
      $\pi: \mathcal{G}\rightarrow [0,1]$ & propensity score function \\
      $\ell: \mathbb{R}\times\mathbb{R} \rightarrow \mathbb{R}^{\geq 0}$ & regression loss function \\
      $c: \{0,1\}\times[0,1]\rightarrow \mathbb{R}^{\geq 0}$ & classification loss function \\
      \hline
    \end{tabular}
    \label{tab:notation}
\end{table}

\section{Problem Setting} 

We assume a training dataset $\mathcal{D}^\text{train} = \{(G_i,y_i){\}_{i=1}^N}$ that includes $N$ molecular graphs, where $G_i \in \mathcal{G}$ is a molecular graph (biasedly) sampled from the universe of molecules $\mathcal{G}$, and $y_i \in  \mathbb{R}$ is the target chemical property.
Each molecular graph $G_i=(\mathcal{V}_i,\mathcal{E}_i, \sigma_i) \in \mathcal{G}$ has a set of nodes (i.e., atoms) $\mathcal{V}_i$, a set of edges (i.e., bonds) $\mathcal{E}_i \subseteq \mathcal{V}_i \times \mathcal{V}_i$, and a label function $\sigma: \mathcal{V}_i \cup \mathcal{E}_i \rightarrow \mathcal{L}$ . Here, $\mathcal{L}$ is the set of possible node labels (i.e., atom types such as hydrogen and oxygen) and edge labels (i.e., bond types such as double bond and aromatic bond).
Our aim is to obtain a prediction function $f: \mathcal{G}\rightarrow \mathbb{R}$ that predicts a particular target chemical property over the molecule universe $\mathcal{G}$ that we are interested in.
We assume that we have a uniformly randomly sampled part of (or possibly the entire) chemical universe, consisting of $M$ molecules $\mathcal{D}^\text{test} = \{G_i{\}}_{i=N+1}^{N+M}$.

This problem is difficult because the training data are constructed from past experiments reported in the literature; therefore, they are significantly biased with respect to the uniform distribution over the chemical universe because of the decisions implemented by researchers on experimental plans or publication options. 
Hence, there is no guarantee that the predictor derived from the biased training data has high predictive performance even on the chemical universe.

Note that the $\mathcal{D}^\text{test}$ can also be a biased sample; however, without loss of generality, we only assume it is a uniformly random subset of the molecules that we are interested in.

In summary, the inputs and outputs of the problem are as follows:
\\
\textbf{Input:} molecular graph $G_i=(\mathcal{V}_i,\mathcal{E}_i, \sigma_i)\in \mathcal{G}$ in $\mathcal{D}^\text{train}$ and $\mathcal{D}^\text{test}$ for training and test, respectively.
\\
\textbf{Output:} one target chemical property $y_i \in \mathbb{R}$.
\\
The ordinary setting is included in our problem setting when $\mathcal{D}^\text{train}$ and  $\mathcal{D}^\text{test}$ come from the same distribution.

We summarize the symbols used in this paper in Table~{\ref{tab:notation}}.

\section{Methods} 
We begin by reviewing the GNN architecture that is the fundamental building block of our model, and then we describe two bias canceling schemes: 
IPS
and 
CFR
. 
They are combined to solve the problem of chemical graph property prediction under experimental biases.


\subsection{GNNs to predict chemical properties} \label{sec:graphconvolution}

Among many successful GNNs, we selected the message-passing GNN architecture proposed by Gilmer et al.~\cite{gilmer2017neural} owing to its generality, simplicity, and fair performance in the chemical domain.

A GNN uses a graph $G=({\mathcal V},{\mathcal E}, \sigma) \in \mathcal{G}$ as its input.
In the $t$-th layer of the GNN, 
it updates the current set of the node representation vectors $\{\mathbf{h}^t_v\}_{v \in V_i}$ to $\{\mathbf{h}^{t+1}_v\}_{v \in V_i}$.
Specifically, the representation vector of node $v$ is updated depending on the current vectors of its neighbor nodes using the following update formula:
\begin{align}
        \mathbf{m}_v^{t+1}&=a \left(\sum_{u \in \mathcal{N}(v)} m_t \left(\mathbf{h}_v^t,\mathbf{h}_u^t,\sigma(v,u)\right)\right), {~~~} \nonumber  \\
        \mathbf{h}_v^{t+1}&=u_t\left(\mathbf{h}_v^t,\mathbf{m}_v^{t+1}\right), \nonumber 
\end{align}
where $\mathcal{N}(v)$ denotes the set of the neighbor nodes of $v$, and $m_t$ is a so-called message passing function that collects the information (i.e., the representation vectors) of the neighbors,
and it is a linear function of $(\mathbf{h}_v^t,\mathbf{h}_u^t)$ depending on the edge label $\sigma(v,u)$.
Further, $a$ is an activation function, for which we select the rectified linear unit (ReLU) function, and $u_t$ is the vertex update function, for which we use the gated recurrent unit (GRU).
The initial node representations $\{\mathbf{h}_v^0\}_{v \in \mathcal{V}}$ are initialized depending on their atom types.  As the message passing operation is repeatedly applied, the node representation vector gradually incorporates information about its surrounding structure.

After being processed through $T$ layers, the final node representations $\{\mathbf{h}_v^{T}\}_{v \in \mathcal{V}}$ are obtained;
they are aggregated to the graph-level representation $\mathbf{h}_{G}$ using a readout function 
   $\mathbf{h}_{G} =r \left(\{\mathbf{h}_v^{T}\}_{v \in \mathcal{V}}\right)$, 
where we could simply use summation, followed by a linear transformation as the readout function. However, in our implementation, we used a slightly more complex solution: a long short-term memory (LSTM) pooling layer, followed by a linear layer.
The graph-level representation $\mathbf{h}_{G}$ is passed to the final layer to achieve the outputs of the GNN, such as the chemical property prediction, propensity score, and domain classification, as we discuss later.

\begin{figure}[t]
    \centering
    \includegraphics[width=.5\linewidth]{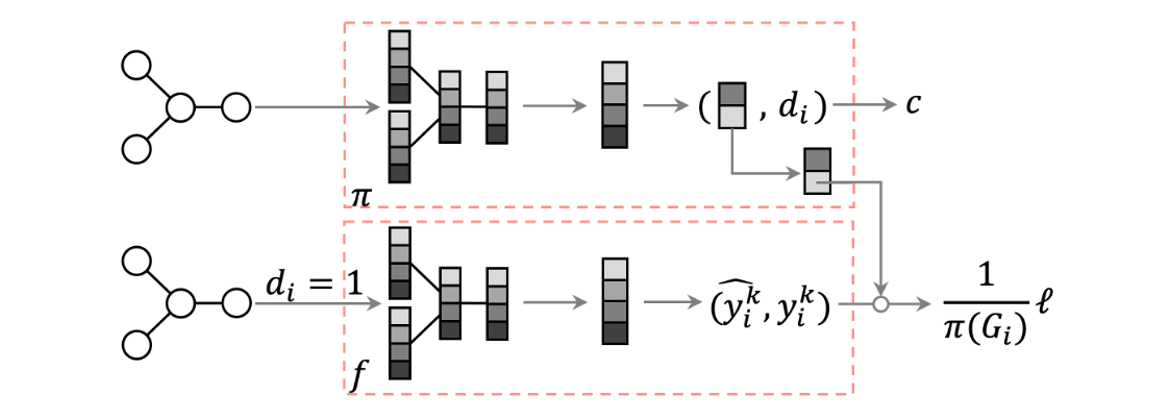}
    \caption{Architecture of the two-step IPS approach.}
    \label{fig:IPS}
\end{figure}
\begin{figure}[t]
    \centering
    \includegraphics[width=.5\linewidth]{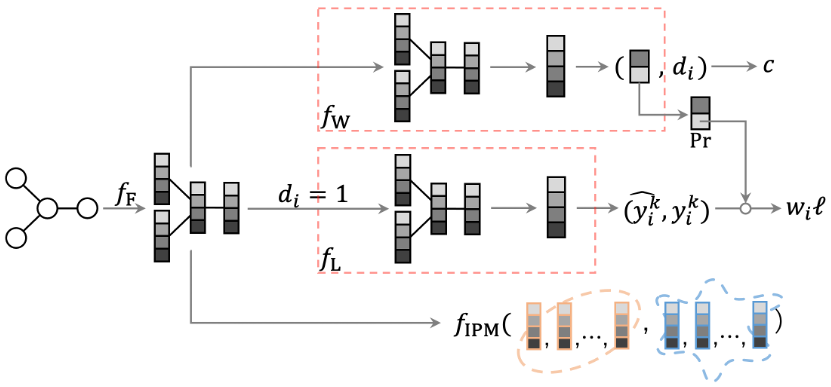}
    \caption{Architecture of the end-to-end CFR approach.}
    \label{fig:CFR}
\end{figure}

\subsection{
Bias correction using 
IPS
}\label{sec:IW}

If we assume that no biases exist in our training dataset, the distributions of the training and target (test) datasets are identical. This implies that minimizing the empirical mean of the loss function $\ell$, that is, 
    $\frac{1}{N}\sum_{i=1}^N \ell(y_i,f(G_i)),$
directly obtains a good prediction model that achieves a small expected loss $\text{E}[\ell(y,g(G))]$ for the test data. 
However, because the training dataset is sampled in a biased manner in our scenario, the minimization of the standard empirical loss results in a biased prediction model.

A possible remedy to this problem is the use of a propensity score~\cite{imbens2015causal} to adjust the importance weight of each training instance.
The propensity score $\pi(G)$ of molecular graph $G$ is the probability that the molecule is included in the experimental data.
The loss for each molecule is inversely weighted with the propensity score, resulting in the modified objective function:
\begin{align} \label{eq:IPSLoss}
    o^\text{IPS}(f)=\frac{1}{N}\sum_{i=1}^N \frac{1}{\pi(G_i)}\ell(y_i,f(G_i)). \nonumber
\end{align}
With the correct propensity score function $\pi$, the weighted loss function is unbiased with respect to the uniform sampling from the molecular universe.
As shown in Figure~\ref{fig:IPS}, 
the IPS approach involves two steps: propensity score estimation and chemical property prediction.
We first estimated the propensity score function that represents the probability of each molecule being experimented.
It is estimated from the biased training dataset and the unbiased test set (uniformly sampled from the molecule universe).
This step is frequently performed by solving a two-class probabilistic classification problem to classify the two datasets using the logistic loss (also called the cross-entropy loss).
Note that the target property values are not used for propensity modeling.
The second step estimates the chemical property prediction model with the loss function weighted using the inverse of the propensity score. We used the squared loss function as $\ell$, and the problem is cast as a weighted regression problem.
The propensity score function and chemical property prediction model are both implemented as GNNs because they use graph-structured molecules as their inputs.


\subsection{
Bias correction using 
CFR
}\label{sec:CFR}
The CFR approach is another option to correct sample selection biases.
We adopted the method proposed by Hassanpour et al~\cite{hassanpour2019counterfactual}, which is an improvement of the best-known method proposed by Shalit et al.~\cite{shalit2017estimating}.
To apply CFR on our tasks, we introduce $d_i\in\{0,1\}$ to indicate the treatment (i.e., domain in general) that $G_i$ belongs to (i.e., test or training). 
The method has four components: a feature extractor, label predictor (replacement of the original treatment outcome predictors), internal probability metric, and weight estimator, denoted by $f_{\text F}$, $f_{\text L}$, $f_{\text {IPM}}$, and $f_{\text W}$, respectively.
Different from the original causal effects estimation setting, only the molecular graphs from $\mathcal{D}^\text{train}$ will be passed trough the treatment outcome predictors because molecular graphs from $\mathcal{D}^\text{test}$ have no labels. Thus, there is only one treatment outcome predictor exists in our model for predicting chemical property, which we call label predictor.
As shown in Figure~\ref{fig:CFR},
in contrast with IPS, which has two separate GNN models, three paths exist in one single model; the first path is for predicting chemical property of $G_i\in \mathcal{D}^\text{train}$, and it is specified by a label predictor $f_{\text L}$ concatenated after a feature extractor $f_{\text F}$. 
The second path, which takes a batch of molecular graphs $\{G_i\}\subseteq\mathcal{D}^\text{train} \cup \mathcal{D}^\text{test}$ as input, is for measuring distance between distributions, which is specified by internal probability metric $f_{\text {IPM}}$ concatenated after the feature extractor $f_{\text F}$.
The third path is for estimating weight of $G_i\in \mathcal{D}^\text{train} \cup \mathcal{D}^\text{test}$, which is specified by an estimator $f_{\text W}$ concatenated after the feature extractor $f_{\text F}$; 
Note that $f_{\text F}$ commonly appears in all of the paths.
The final deliverable $f$ that we desire is the composite function of $f_{\text F}$ and $f_{\text L}$; that is, $f(G)=f_{\text L}(f_{\text F}(G)$.
The weight estimator and the internal probability metric are not our final objective, but they aid in extracting debiased representations of the inputs in the training phase.

In the training phase, the first path (consisting of $f_{\text F}$ and $f_{\text L}$) aims to predict the target chemical property, by minimizing
\begin{equation}
    o^\text{property}(f_{\text F},f_{\text L}) = \frac{1}{N} \sum_{i=1}^N w_i \ell(y_i, f_{\text L}(f_{\text F}(G_i))), \nonumber
\end{equation}
where $\ell$ is the loss function for chemical properties, namely, the squared loss in our case. Term $w_i$ is the importance sampling weight.
According to Hassanpour et al~\cite{hassanpour2019counterfactual}, if we denote $\phi_i=f_{\text W}(f_{\text F}(G_i))\in\mathbb{R}^2$ (outputs of two domains) and use softmax function to obtain probabilities, $w_i=\frac{1}{\Pr(t_i|G_i)}=1+e^{\phi_i^{(1-t_i)}-\phi_i^{(t_i)}}$ when $N=M$, which is similar to inverse propensity score in the IPS approach.
Note that in our case, $t_i$ is fixed to $1$ while calculating $w_i$ because only $G_i\in \mathcal{D}^\text{train}$ will be passed through $f_{\text L}$.
In the second path (including $f_{\text F}$ and $f_{\text {IPM}}$), the internal probability metric $f_{\text {IPM}}$ aims to measure the distance between distributions of molecular graphs from $\mathcal{D}^\text{train}$ and $\mathcal{D}^\text{test}$.
The objective function for the second path is expressed as
\begin{equation}
    o^\text{IPM}(f_{\text F},f_{\text {IPM}}) = f_{\text {IPM}}(\{f_{\text F}(G_i)\}_{i:d_i=0}, \{f_{\text F}(G_i)\}_{i:d_i=1}).
    \nonumber
\end{equation}
In the third path (including $f_{\text F}$ and $f_{\text W}$), the $f_{\text W}$ aims to correctly classify the domain (i.e., test or training) of the input.
Denoting the loss function for domain classification (i.e., the cross-entropy loss in our case) by $c$, the objective function for the third path is expressed as
\begin{align}
       o^\text{domain}(f_{\text F},f_{\text W}) = \frac{1}{N+M} \sum_{i=1}^{N+M} c(d_i, f_{\text W}(f_{\text F}(G_i))).  \nonumber
\end{align} 
Parameters of $f_{\text F}$, $f_{\text L}$, and $f_{\text {IPM}}$ are updated by minimizing the objective function $o(f_\text{F},f_\text{L},f_\text{IPM})=o^\text{property}(f_{\text F},f_{\text L}) + \alpha \cdot o^\text{IPM}(f_{\text F},f_{\text {IPM}})$, where $\alpha$ is a hyper-parameter.
After the update of $f_{\text F}$, $f_{\text L}$, and $f_{\text {IPM}}$, parameters of $f_{\text W}$ are then updated by minimizing $o^\text{domain}(f_{\text F},f_{\text W})$.

\section{Experiments} \label{sec:experiments}

\subsection{Biased sampling scenarios} \label{sec:Biased sampling scenarios}
Because we cannot know why the compounds reported in the literature were selected, we simulated several possible scenarios to sample our training datasets.
We considered the following four possible biased sampling scenarios:
\begin{itemize}
\item Scenario 1 assumes that molecules with a smaller number of atoms have higher sampling chances, because smaller molecules are considered more common and better explored~\cite{lipinski2004lead}.
\item Scenario 2 prefers molecules with smaller proportions of single bonds (i.e., with higher proportions of the other bond types). The spectral manifestations of chemical functional groups significantly depend on bond types within the group~\cite{C8CP07057A}. For simplicity, we assumed that less single-bonded molecules would have stronger spectral manifestations.
\item Scenario 3  selects molecules with higher values of the gap between the highest energy occupied molecular orbital (HOMO) and lowest energy unoccupied molecular orbital (LUMO), because larger HOMO-LUMO gap values often indicate higher stability and lower chemical reactivity~\cite{aihara1999reduced}. 
\item Scenario 4  assumes that scientists focus on compounds with high target property values based on their expertise and experience to conduct experiments. Further, compounds with higher targets values have higher sampling chances.
\end{itemize}

\subsection{Datasets} \label{sec:Datasets}
We used the chemical molecule dataset QM9~\cite{ramakrishnan2014quantum} as the universe $\mathcal{G}$ of the small graphs.
QM9 is a publicly available dataset containing 134,000 small stable organic molecules composed of hydrogen (H), carbon (C), oxygen (O), nitrogen (N), and fluorine (F).
They are the subset of all the molecules with up to nine heavy atoms (C, O, N, F) out of the 166 billion molecules of the GDB-17 chemical universe~\cite{ruddigkeit2012enumeration,ramakrishnan2014quantum}; therefore,  the QM9 dataset can be considered to have a close distribution to the natural universe of molecules comprising C, O, N, and F.
Each molecule in QM9 has 12 fundamental properties~\cite{ramakrishnan2014quantum}: dipole moment (mu), isotropic polarizability (alpha), HOMO (homo), LUMO (lumo), gap between the homo and lumo (gap), electronic spatial extent (r2), zero point vibrational energy (zpve), internal energy at 0 K (u0), internal energy at 298.15 K (u298), enthalpy at 298.15 K (h298), free energy at 298.15 K (g298), and heat capacity at 298.15 K (cv); they were used as 12 targets for 12 regression tasks.

We further used three chemical molecule datasets, i.e., ZINC~\cite{gomez2018automatic}, ESOL, and FreeSolv~\cite{wu2018moleculenet}.
The ZINC dataset contains 250,000 drug-like commercially available molecules graphs with up to 38 heavy atoms. The task of ZINC is to predict a molecular property known as constrained solubility.
ESOL is a small dataset consisting of water solubility data for 1,128 molecular graphs.
FreeSolv is also a small dataset consisting of 643 molecular graphs with hydration free energy of small molecules in water.
Note that molecule structures in these three datasets are far less than exhaustively enumerated. 
Thus, different from the QM9 dataset, there is no guarantee that the $\mathcal{D}^\text{test}$ of these three datasets can be regarded as an unbiased subset of the molecular universe. 
However, without loss of generality, we only assume the $\mathcal{D}^\text{test}$ is unbiased or less biased (compared with the $\mathcal{D}^\text{train}$) from the indicators that we used to induce biases.

\begin{figure*}[t]
    \centering
    \includegraphics[width=\textwidth]{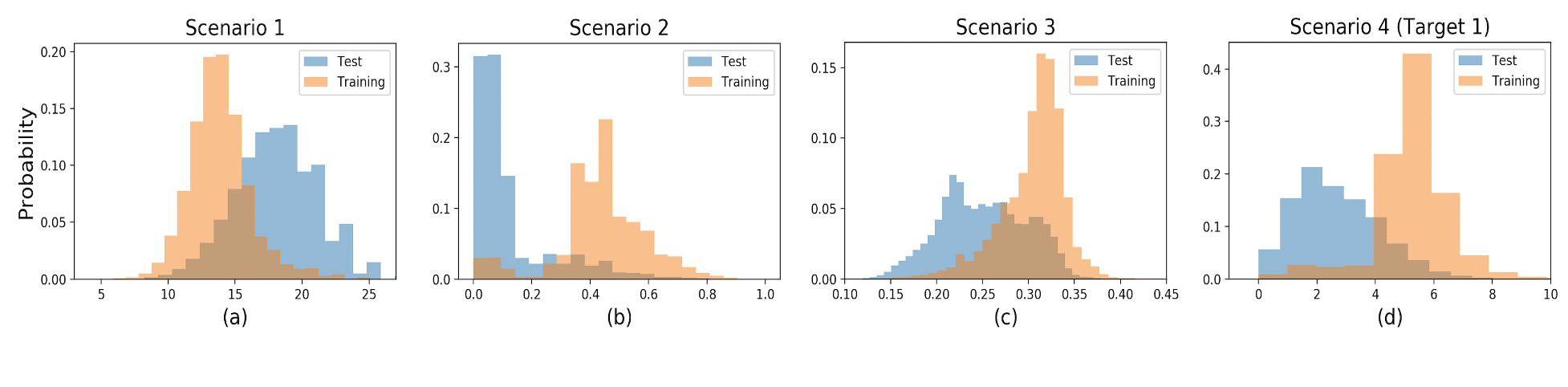}
    \caption{Average densities of the training and test datasets of QM9.
    The $x$-axises correspond to (a) number of atoms, (b) proportion of double/triple/aromatic bonds, (c) value of HOMO-LUMO gap (gap), and (d) value of mu, respectively.
    }
    \label{fig:Densities}
\end{figure*}

We first sampled a test dataset with a size of 10\% of the entire dataset uniformly at random.
Then, according to each of the four biased sampling scenarios, we sampled a biased training dataset 
from the remaining molecules.
Each compound had a sampling chance determined by the sigmoid function depending on the corresponding sampling criteria.
Take QM9 for example, in Scenario 1, the smallest molecules with three atoms had the largest sampling chances, while the ones with 27 atoms had the smallest chances.
The larger the gain of the sigmoid curve becomes, the more the training and test datasets were separated; we tuned the gain to bring the average sampling probability to 10\%.
We used sampling without replacement; therefore, no graph belonged to the training and test sets simultaneously.
We repeated the sampling procedure 30 times to build  training and test datasets for statistical testing.
Figure~\ref{fig:Densities} shows the average densities of the 30 trials under the biased sampling scenarios of QM9 (for Scenario 4, only the first target among the 12 targets is shown because of the page limitation).

\subsection{Implementation details} \label{sec:Implementation details}

We used the PyTorch Geometric library~\cite{Fey/Lenssen/2019} to implement the GNN models.  
In QM9 and ZINC, each atom is encoded as a 13 and 28-dimensional vector (one hot), respectively, depending on the atom type.
In ESOL and FreeSolv, we followed the settings in MoleculeNet~\mbox{\cite{wu2018moleculenet}}, where each atom is encode as a 9-dimensional vector.
The message passing function $m_t$ depends on edge types and is 32-dimensional. 
The update function $u_t$ is a gated recurrent unit with 32 internal dimensions.
The readout function $r$ is a sequence-to-sequence layer followed by two linear layers with 32 internal dimensions.

The IPS approach required two GNN models, one for the propensity score function and the other for chemical property prediction.
In the former, we used $T=3$ GNN layers and a logistic function as the final layer because the propensity score indicated the probability that a chemical compound is observed.
We used the ADAM~\cite{kingma2014adam} optimizer with no weight decay and a batch size of 64 for each iteration.
We used a learning rate updating scheduler with an initial learning rate of $1e-5$; this reduced the learning rate by a factor of 0.7 until the validation error stopped reducing for five training epochs.
The validation datasets were randomly selected to include 20\% of the training and test sets.
The optimized model that achieved the lowest validation error on the validation dataset was applied to infer the importance.
The chemical property prediction models also had $T=3$ GNN layers, and the other training settings were almost the same as those for the propensity score model. The validation set was 20\% of the training set.
Because we had 15 target chemical properties, we trained 15 different GNN predictors.

The network structure for the CFR approach had three output paths: the label predictor, internal probability metric, and weight estimator. The label predictor and weight estimator shared the common feature extraction layers on the input side.
We set the number of the GNN layers corresponding to the feature extractor and those for the weight estimator and label predictor to $3$.
Similar to the IPS approach, the weight estimator had a readout function for classification  and the label predictor had one for regression. Note that the feature extractor had no readout function.
We used Wasserstein distances~\cite{cuturi2014fast} for the internal probability metric.
In addition, before the internal probability metric, there was readout function to aggregate features extracted by the feature extractor to a batch of graph-level features.
We set the $\alpha$ for QM9, ZINC, ESOL, and FreeSolv to 10, 10, 100, and 10, respectively.
As with the IPS approach, we used ADAM~\cite{kingma2014adam} with no weight decay as the optimizer. 
We also used a learning rate updating scheduler, with an initial learning rate of $1e-5$ and reduced the learning rate by a factor of 0.7 until the validation error stopped reducing for five training epochs.
The batch size was set to 64.
In contrast with IPS, the entire network was trained in an end-to-end manner.
We trained 15 GNNs for each of the target properties in each trial.
20\% of the training and test sets were used to validate the domain classifier, and 20\% of the training set was used for the label predictor.

As the baseline method, we used the same GNN structure as the one for IPS, but without bias mitigation.
In contrast to the IPS approach, we used the standard unweighted average loss for the training dataset.
The same settings as for IPS were used except for those specific to IPS, such as the number of GNN layers, the selection of the hyperparameters, and the training and validation sets.

The prediction accuracy for each option was evaluated in terms of the average mean absolute error (MAE) obtained from the 30 trials.
We also performed the paired $t$-test to check the statistical significance of performance differences.

\begin{figure*}[t]
    \centering
    \includegraphics[width=\textwidth]{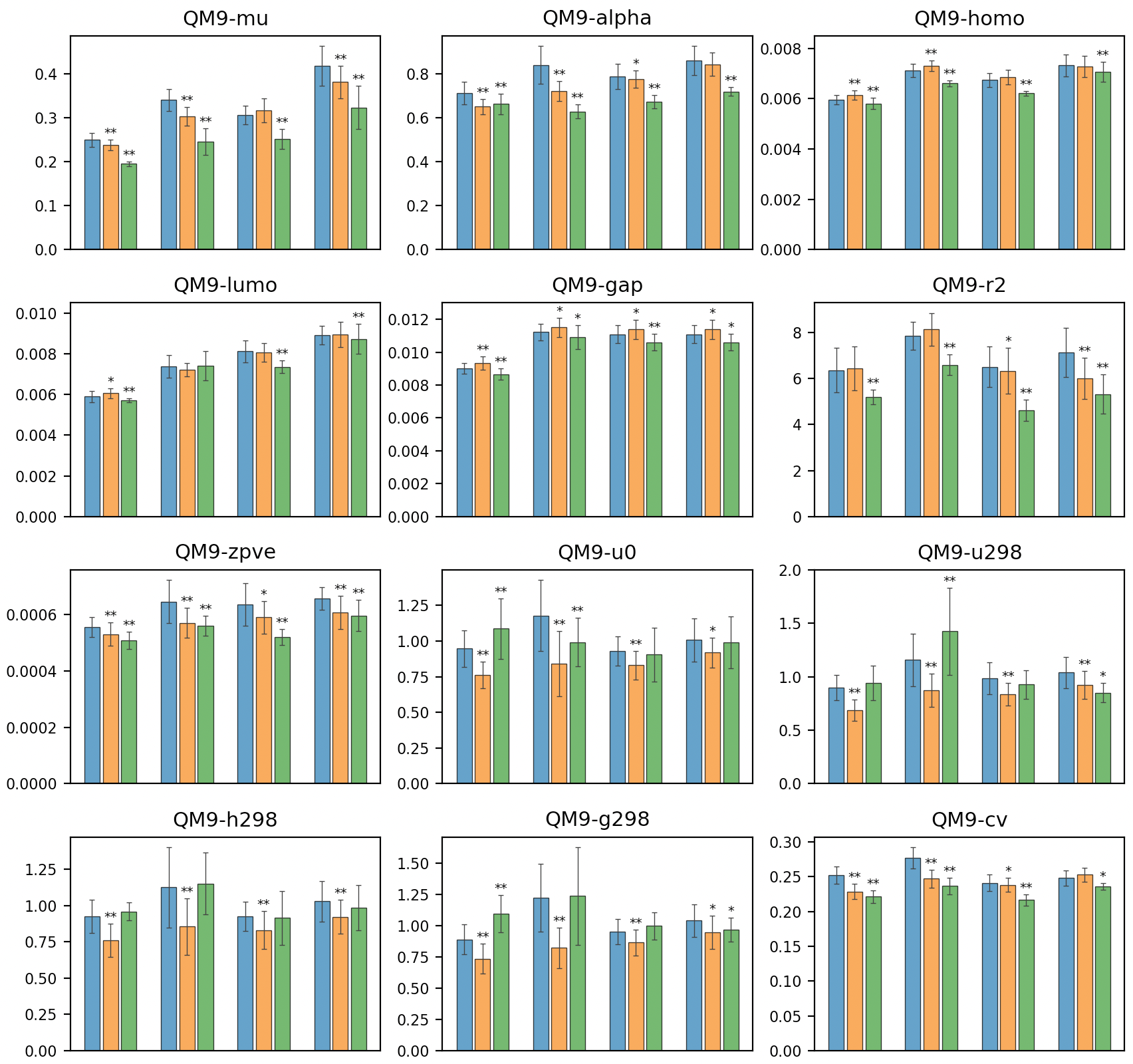}
    \caption{MAE comparison results of QM9. The 12 subplots correspond to 12 property prediction tasks. The x-axises correspond to the four simulated biased sampling scenarios. The blue, orange, and green bars correspond to the Baseline, IPS, and CFR approaches, respectively. 
    }
    \label{fig:mae1}
\end{figure*}

\begin{figure*}[t]
    \centering
    \includegraphics[width=.9\textwidth]{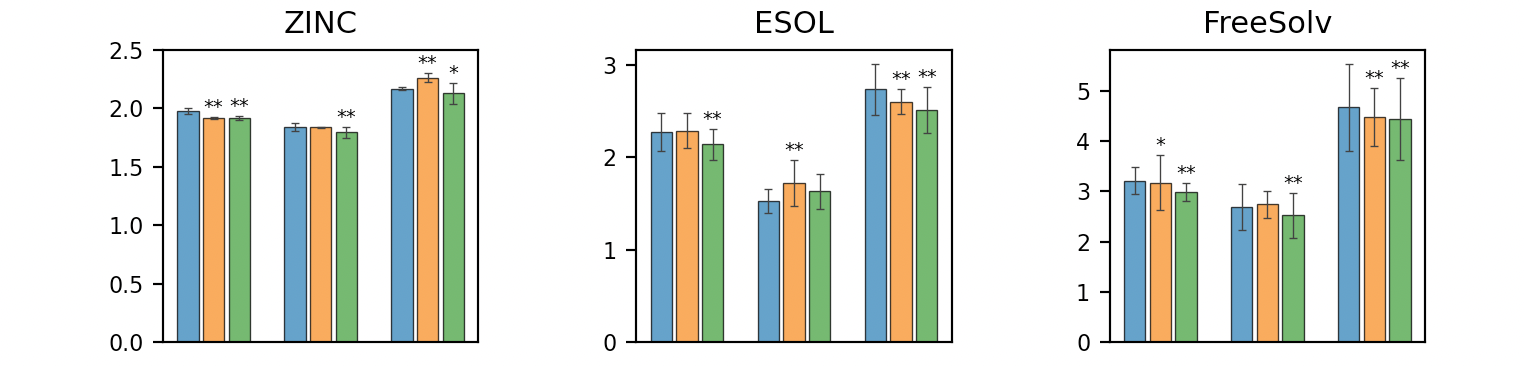}
    \caption{MAE comparison results of ZINC, ESOL, and FreeSolv. The x-axises correspond to the three simulated biased sampling scenarios (Scenario 3 was not performed on them). The blue, orange, and green bars correspond to the Baseline, IPS, and CFR approaches, respectively. 
    }
    \label{fig:mae2}
\end{figure*}

\subsection{Results}

\subsubsection{Comparison of predictive performance.}
Figures~\ref{fig:mae1} and Figures~\ref{fig:mae2}
show all the MAE comparison results (in the mean and standard deviations of 30 trials) corresponding to the four simulated biased sampling scenarios. 
The * symbols above the bars denote the $p$-values of the paired $t$-test when comparing Baseline (without bias mitigation) with IPS and CFR, respectively; ** means the $p$-value was less than 0.01, and * means that it was less than 0.05.
Note that, in the case of QM9, the result for the HOMO-LUMO gap (denoted by `gap') in Scenario 3 was
equivalent to that in Scenario 4 because QM9 contains HOMO-LUMO gap information, 

The overall results for all the scenarios indicated that the IPS approach improved the performance for many properties and scenarios; in particular the performance was statistically significantly improved for the six properties of QM9 (alpha, zvpe, u0, u298, h298, and g298) in all of the four scenarios, which indicated that IPS has solid effectiveness and potential in mitigating experimental biases on these tasks. 
However, we found that there were some statistically insignificant comparison and even significant failure for the three properties of QM9 (homo, lumo, gap) and the properties of ZINC, ESOL, and FreeSolv.
These failures indicated that although IPS achieved improvements on most of the tasks, it was not stable.
In addition, the improvements by IPS of QM9 were more significant for Scenarios 1 and 2 than Scenarios 3 and 4.
The differences might be explained by the accuracy of the propensity score model; the accuracies in the four scenarios were 81.05\%,  87.49\%, 76.04\%, and 79.02\%, respectively, which meant that the propensity scores were more accurate in Scenarios 1 and 2.

The CFR approach achieved more remarkable predictive performance than the IPS approach for most of the properties and scenarios.
For the three properties of QM9 (homo, lumo, gap) and the three properties of ZINC, ESOL, and FreeSolv in all of the four scenarios, where IPS failed to improve the predictive performance, CFR achieved statistically significant improvements comparing to the baseline method.
Further, for the two properties of QM9 (alpha, zvpe), the improvements achieved by CFR were more remarkable than IPS.
However, we found that for the four properties of QM9 (u0, u298, h298, g298), where IPS achieved remarkable improvements, CFR failed to show statistically significant increasing or decreasing of predictive performance comparing to the baseline method.

In summary, both of the IPS and CFR approaches made solid improvements in mitigating experimental biases for most of the properties and scenarios, and CFR showed better performance than IPS.

\begin{figure*}[t]
    \centering
    \includegraphics[width=\textwidth]{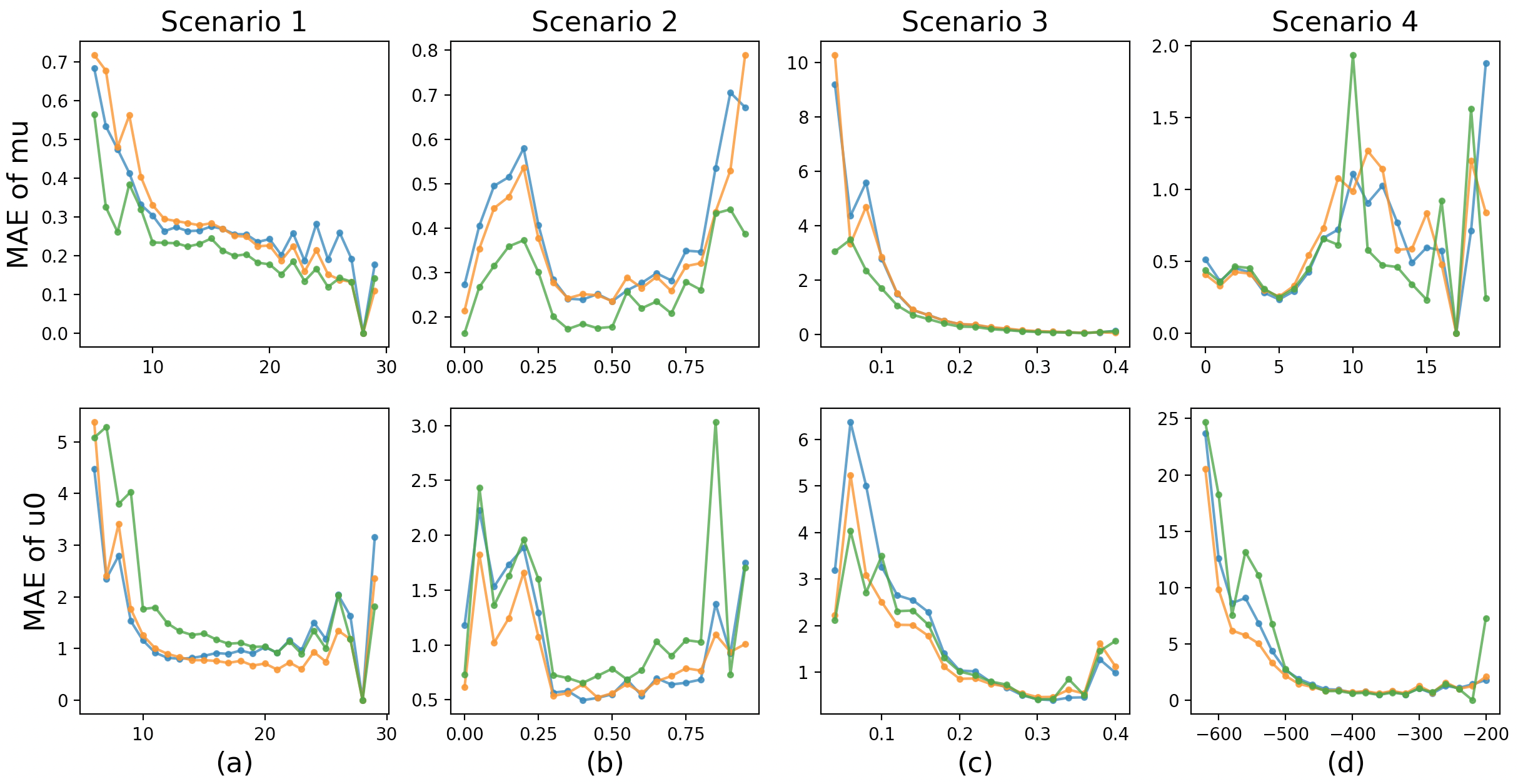}
    \caption{MAE comparison results of mu and u0 depending on indicators of the four biased sampling scenarios. The $x$-axises correspond to (a) number of atoms, (b) proportion of double/triple/aromatic bonds, (c) value of HOMO-LUMO gap (gap), and (d) value of the target (i.e., mu for the top and u0 for the bottom), respectively. The blue, orange and green lines correspond to the Baseline, IPS, and CFR approaches, respectively.
    }
    \label{fig:mae_indicators}
\end{figure*}

\subsubsection{Prediction accuracy depending on indicators for biased sampling.}
We further investigated why the IPS and CFR approaches successfully corrected the bias by visualizing the prediction accuracy depending on the indicators used in the biased sampling scenarios.
Because of space limitations, we only show the predictive performance for the chemical property mu and u0 of QM9 in Figure~\ref{fig:mae_indicators}. 
Similar results can be observed on other tasks.
The horizontal axes in the figure correspond to the indicators, namely, (a) the number of atoms, (b) the proportion of double/triple/aromatic bonds, (c) the value of the HOMO-LUMO gap (gap), and (d) the value of the target (i.e., mu for the top and u0 for the bottom), respectively.
The top four figures correspond to the average test MAE (that is, the smaller, the more accurate) of predicting mu of QM9 and the bottom four figures correspond to the average test MAE of predicting u0 of QM9.

For predicting mu, according to the MAE comparison results, we know that both IPS and CFR achieved significant improvements of predictive performance for all scenarios and CFR outperformed IPS.
Under Scenario 1, most of the molecules used for training had a number of atoms ranging from 5 to 15.
For predicting mu of molecules with 15 or more atoms, primarily in the test dataset, IPS and CFR consistently outperformed the baseline method, and in addition, CFR consistently outperformed IPS.
Under Scenario 2, most of the molecules used for training had a proportion of double/triple/aromatic bonds higher than 0.4.
Again, on those molecules with proportion less than 0.4, IPS and CFR consistently performed better than the baseline method, and CFR almost outperformed IPS.
Similarly in Scenarios 3 for predicting mu, we observed the advantage of CFR for the smaller indicator values (less than 0.2) corresponding to the test datasets.
However, we failed to observe the similar advantage of CFR obviously in Scenario 4 for predicting mu.

For predicting u0, we know that IPS achieved significant improvements of predictive performance for all scenarios while CFR failed.
For predicting mu of molecules with 15 or more atoms, IPS consistently outperformed the baseline method while CFR almost failed.
Under Scenario2, both IPS and CFR achieved better performance on those molecules with proportion less than 0.4, and IPS showed better performance, which was consistent with our findings.
Similar in Scenario 3 and 4 for predicting u0, we observed the advantage of IPS for the smaller indicator values corresponding to the test datasets while we failed to observe the advantage of CFR.

In summary, by visualizing the prediction accuracy depending on those indicators for biased sampling, we can partially conclude that, on most of the tasks, IPS and CFR improved the predictive performance on molecules with lower chance for observation, which led to the overall improvements.

\subsection{Future scope}

The fundamental direction of our research is to mitigate experimental biases in scientific research.
In this study, we only focused on the population of chemical compounds and the task of property prediction.
Possible future topics are extensions to more complex prediction tasks under experimental biases, including chemical-chemical link prediction in the chemical domain, biological network prediction in the biological domain, and discovery of unknown compounds in the material science domain.
Another possible direction is applications of more modern methods in  covariate shift, causal inference, and domain adaptation.
Since the difficulty in our problem setting lies mainly in learning unbiased representation of biased observed instances, it would be promising to adapt general ideas from those related areas and to develop theories and techniques for our specific tasks.

\section{Conclusion} \label{sec:conclusion}
We considered the prediction of chemical properties from datasets that have experimental biases.
We introduced two promising bias mitigation techniques by combining the recent developments in causal inference and GNN-based graph learning.
We tested four practical biased sampling scenarios on the well-known QM9, ZINC, ESOL, and FreeSolv datasets for experiments.
The experimental results confirmed that the two approaches improved the predictive performance in all scenarios on most of the tasks with statistical significance compared with the baseline method, which had no effort for bias mitigation.
We also found that the more modern CFR approach outperformed the IPS approach for most of the tasks and scenarios.

\bibliographystyle{splncs04}
\bibliography{lncs}

\end{document}